\documentclass[a4paper,11pt]{article}
\usepackage{pos}
\usepackage{bm}% bold math
\usepackage{braket}

\title{Recent progress on charmed hadron interactions from lattice QCD}
\author*[a]{Yan Lyu}
\affiliation[a]{RIKEN Center for Interdisciplinary Theoretical and Mathematical Sciences (iTHEMS), RIKEN, \\
Wako 351-0198, Japan}

\emailAdd{yan.lyu@riken.jp}

\abstract{
Recent years have witnessed rapid progress in charmed hadron physics, driven by numerous experimental discoveries of exotic states such as $T^+_{cc}$ and $P_c$.
These findings have highlighted the importance of understanding charmed hadron interactions.
With significant advances in theory and computation, lattice QCD has become a reliable tool for studying nonperturbative dynamics of low-energy QCD.
In this talk, I review recent lattice QCD studies of charmed hadron interactions, focusing on three representative systems: $D^*$-$D$, $N$-$c\bar c$, and $\Omega_{ccc}$-$\Omega_{ccc}$.}

\FullConference{The 21st International Conference on Hadron Spectroscopy and Structure (HADRON2025)\\
 27 - 31 March, 2025\\
Osaka University, Japan\\}

\begin{document}
\maketitle

\section{Introduction}
Hadron-hadron interactions govern the properties of few- and many-body quantum systems across a wide range of scales, from exotic hadronic states~\cite{Chen2022_PPP,Guo2017} 
and finite atomic nuclei~\cite{Shen:2019dls} to dense nuclear matter relevant to neutron stars~\cite{Hayano-Hatsuda2010,Baym:2017whm}. 
At the microscopic level, these interactions emerge as residual forces arising from the strong interactions among quarks and gluons, and are thus ultimately dictated by Quantum Chromodynamics (QCD).

The nucleon-nucleon interaction has been extensively studied both experimentally and theoretically for decades. 
These efforts are well summarized in a number of high-precision models of nuclear force, which are constrained by thousands of experimental data including nucleon-nucleon scattering and properties of the deuteron bound state. 
Among them are the Argonne v18 potential~\cite{Wiringa1995}, the charge-dependent Bonn potential~\cite{Machleidt:2000ge}, and potentials derived from chiral effective field theory~\cite{Epelbaum:2014sza,Entem:2017gor,Lu:2021gsb}.
In contrast, hadron interactions in the charm quark sector remain much less constrained experimentally. 
This is because that the unstable nature of charmed hadrons makes traditional scattering experiments particularly challenging, and therefore leads to limited available data
\footnote{For recent experimental progress from the scattering and the femtoscopy, see plenary talks by Dr. Koji Miwa and Dr. Laura Fabbietti in this conference.}. 
Nevertheless, understanding such interactions is crucial, as they not only serve as important input for studying charmed exotic hadrons and charmed hypernuclei, 
but also give access to distinctive phenomena that are either absent or less prominent in light-flavor systems, such as the appearance of the heavy quark spin symmetry and enhanced coupled-channel dynamics {\it etc}.

Under these circumstances, deriving charmed hadron interactions from first-principle QCD calculations becomes critically important.
Lattice QCD is the only known systematically improvable approach to nonperturbative calculations of QCD, and has already demonstrated remarkable success in reproducing single-hadron spectra with high precision~\cite{Aoyama:2024cko}.
There are two approaches commonly used to study hadron-hadron interactions from lattice QCD, the finite-volume method~\cite{Luscher1991} and the HAL QCD method~\cite{Ishii2007,Ishii2012}.
The finite-volume method extracts scattering phase shifts by solving a quantization condition that relates the discrete energy spectra of two-hadron systems, typically obtained from {\it temporal} correlation functions, to infinite-volume scattering observables.
In contrast, the HAL QCD method defines an effective potential through {\it spatiotemporal} correlation functions, allowing to calculate various observables with the potential.
A key advantage of the HAL QCD method is that, by leveraging both spatial and temporal information, it extracts hadron-hadron interactions without requiring these correlation functions to be dominated by a single state.
This makes it especially suitable for studying systems involving charmed hadrons under large lattice volumes, 
where the dense two-hadron spectra arising from the heavy charm quark mass and large volumes make isolating individual states challenging.

In this talk, I will review recent development on lattice QCD studies of charmed hadron interactions using the HAL QCD method \footnote{A plenary talk in this conference by Dr. Jozef Dudek reviewed recent progress using the finite-volume method.}.
In particular, I will present three representative examples spanning meson-meson, baryon-meson, and baryon-baryon systems: $D^*$-$D$, $N$-$c\bar c$, and $\Omega_{ccc}$-$\Omega_{ccc}$.

\section{HAL QCD method}
In this section, we briefly review the HAL QCD method~\cite{Ishii2007,Ishii2012}. To proceed, let us consider the $R$-correlator for a system composed of two hadrons $H_1$ with mass $m_1$ and $H_2$ with mass $m_2$,
\begin{align}\label{Eq_R}
    R(\bm r,t)&=\sum_{\bm x}\braket{0|H_{1}(\bm x+\bm r, t)H_{2}(\bm x, t)\overline{\mathcal{J}}(0)|0}/e^{-(m_1+m_2)t}\\ \nonumber
    &=\sum_{n}a_n\psi_{E_n}(\bm r)e^{-(\Delta E_n)t} + O(e^{-(\Delta E^*)t}),
\end{align}
where $\mathcal{J}(0)$ is a source operator, and $a_n=\braket{H_1, H_2; E_n|\overline{\mathcal{J}}(0)|0}$ is the overlapping factor to the $n$-th eigenstate $|E_n; H_1, H_2\rangle$.
$\Delta E_n=E_n-(m_1+m_2)$ is the eigenenergy with respect to the two-hadron threshold, and the inelastic threshold is typically determined by $\Delta E^*\sim m_\pi$ or $\Delta E^*\sim \Lambda_\text{QCD}$.
$\psi_{E_n}(\bm r)$ is the Nambu-Bethe-Salpeter amplitude, defined as,
\begin{align}
    \psi_{E_n}(\bm r) = \sum_{\bm x}\braket{0|H_{1}(\bm x+\bm r, 0)H_{2}(\bm x, 0)| E_n; H_1, H_2}.
\end{align}   

It has been shown in Ref.~\cite{Ishii2012} that $R(\bm r, t)$ satisfies the following integrodifferential equation when $t\gg 1/\Delta E^*$,
\begin{align}
    &\left[\frac{1+3\delta^2}{8\mu}\frac{\partial^2}{\partial t^2}-\frac{\partial}{\partial t} +\frac{\nabla^2}{2\mu} +O(\delta^2\partial^3_t)\right] R(\bm r,t)\\ \nonumber
    &=\int d\bm r' U(\bm r, \bm r')R(\bm r',t), \quad \mu=\frac{m_1m_2}{m_1+m_2}, \quad \delta=\frac{m_1-m_2}{m_1+m_2}.
\end{align}
In practice, the nonlocal potential defined above usually is expanded as $U(\bm r,\bm r')=V(r)\delta(\bm r-\bm r')+\sum_{n=1}V_n(\bm r)\nabla^n\delta(\bm r-\bm r')$.
To describe near-threshold scattering, a local potential at the leading order in the above expansion is sufficiently accurate in most cases, which leads to,
\begin{align}\label{Eq_LO_V}
    V(r)=R^{-1}(\bm r,t)\left[\frac{1+3\delta^2}{8\mu}\frac{\partial^2}{\partial t^2}-\frac{\partial}{\partial t} +\frac{\nabla^2}{2\mu} + O(\delta^2\partial^3_t)\right] R(\bm r,t),
\end{align}
where the $O(\delta^2\partial^3_t)$ term is neglected in our study, as it vanishes exactly for $m_1 = m_2$ and is found to be numerically consistent with zero within statistical uncertainties when $m_1 \neq m_2$.
The truncation error from higher-order terms of the  derivative expansion will be quoted into systematic errors by an estimation through the $t$ dependence of $V(r)$~\cite{Lyu:2025lnd,Lyu:2025ncq}.

\begin{figure}[htbp]
    \centering
    \includegraphics[width=14.0cm]{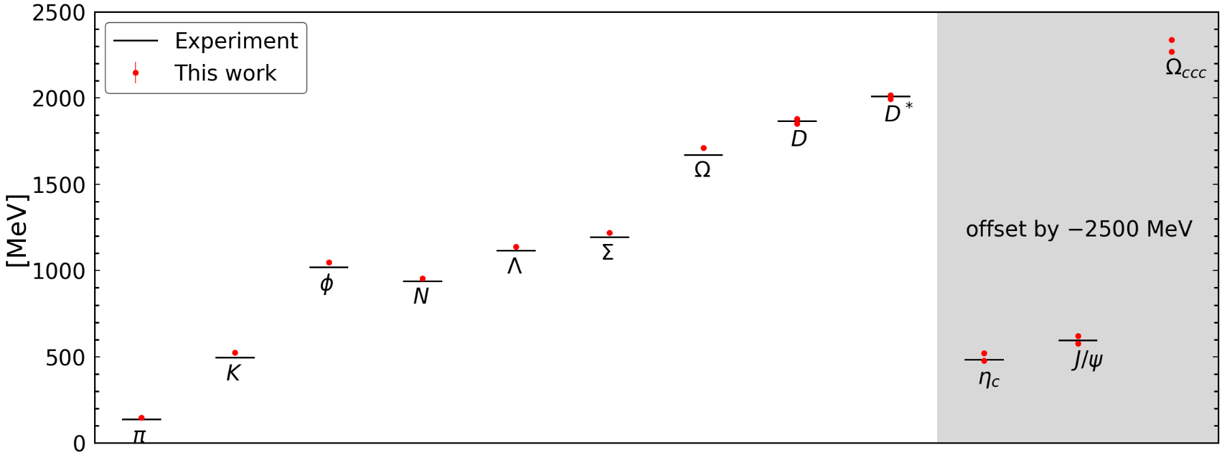}
    \caption{A comparison on single-hadron masses between the experiment and the lattice calculations. The experimental values represented by black lines are from the PDG~\cite{PDG2020}, while the red dots are from our lattice QCD calculations. 
    Two lattice data for each charmed hadron correspond to two sets of parameters in the charmed quark action~\cite{Namekawa2017}, and are linearly extrapolated to the physical charm quark mass.
    The masses for $\eta_c$, $J/\psi$, and $\Omega_{ccc}$ have been shifted vertically for better visibility.
    }
    \label{Fig-hadron-mass}
\end{figure}

Results to be presented in the next few sections are obtained using the ($2+1$)-flavor lattice QCD configurations with almost physical light quark masses 
($m_\pi\simeq146$ MeV and $m_K\simeq525$ MeV) and physical charm quark mass (the mass of the spin-averaged $1S$ charmonium $m_{\bar{c}c}\simeq3068$ MeV)~\cite{Ishikawa2016,Namekawa2017}.
The lattice spacing is $a\simeq0.0846$ fm, and lattice volume is $L^4=96^4$, leading to $La\simeq8.1$ fm, which is large enough to accommodate two hadrons.
Single-hadron masses determined from our lattice calculations are summarized in Fig.~\ref{Fig-hadron-mass}. 
%==========================
%==========================

\section{The $D^*$-$D$ interaction}
The quest of exotic hadrons with multi quark configuration beyond the conventional constituent quark model has been one of central subjects in the study of nonperturbative QCD for decades.
Among others, the doubly charmed tetraquark $T^+_{cc}$ discovered recently by the LHCb collaboration~\cite{LHCb2021_NP,LHCb2021} has attracted special attentions, 
as it represents the long-anticipated state with the genuine tetra-quark configuration due to its doubly charmed nature. 

\begin{figure*}[htbp]
    \centering
    \includegraphics[width=14.0cm]{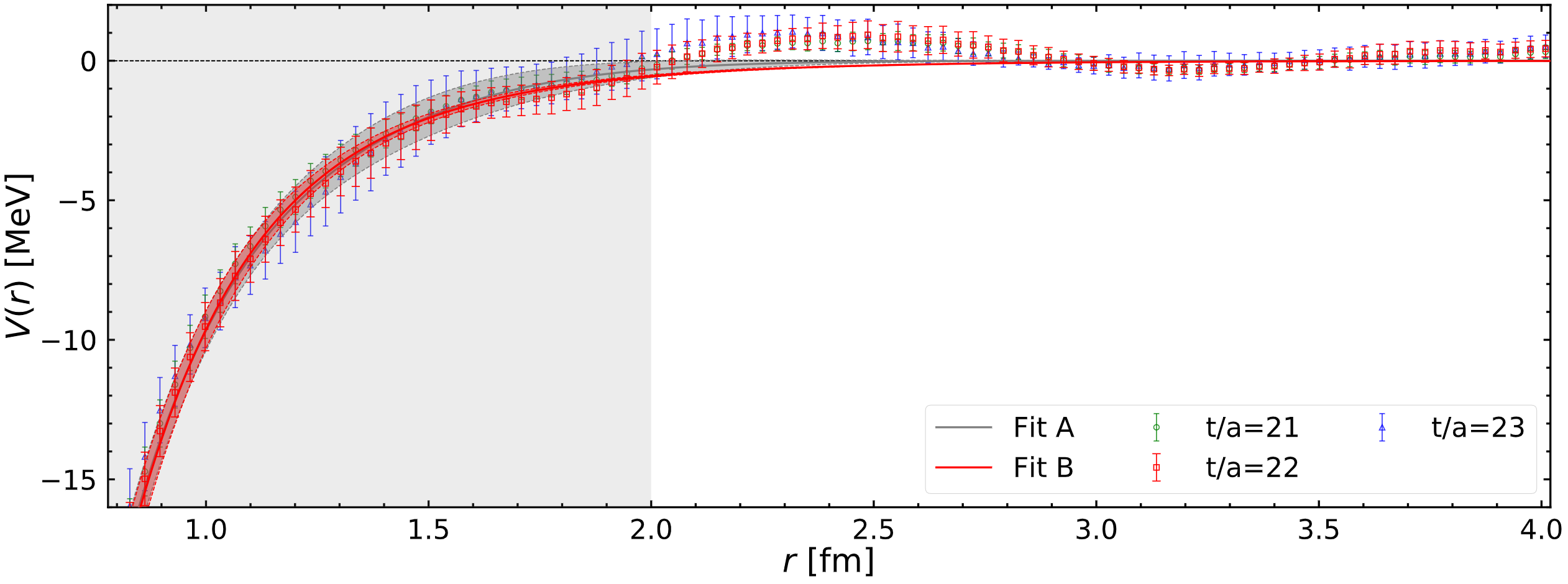}
    \caption{The potential between a $D$ meson and a $D^*$ meson in the $T^+_{cc}$ channel ($I=0, J^P=1^+$). The lattice data for $r<0.8$ fm show monotonically decreasing towards the origin, where $V(r=0)\simeq -500$ MeV.
    The two bands are the fitted results using fit functions in Eqs.~(\ref{Eq_Tcc_fit_A}) and (\ref{Eq_Tcc_fit_B}).
    }
    \label{Fig-Tcc-V}
\end{figure*}

In Fig.~\ref{Fig-Tcc-V}, we show the potential between a $D$ meson and a $D^*$ meson in the $T^+_{cc}$ channel ($I=0, J^P=1^+$)~\cite{Lyu:2023xro}.
The potential is attractive for all distances: the short-range attraction indicates a relation to a diquark configuration $(\bar{u}\bar{d})_{\bm{3}_c, I=J=0}-(cc)_{\bm{3}^*_c, J=1}$ that couples to asymptotic $DD^*$ state, 
in which a large attraction is expected (see \cite{Noh2021} and references therein),
while the long-range potential ($1<r<2$ fm) is found to be consistent with two-pion exchange (TPE) numerically.
We fit the lattice potential by two different fit functions: a phenomenological four-range gaussian ($V^A_\text{fit}$), and a TPE motivated fit ($V^B_\text{fit}$) where  a TPE function is explicitly included to account for the potential tail at long distances. 
\begin{align}
   & V^A_\text{fit}(r) =\sum^4\limits_{i=1}a_ie^{-(r/b_i)^2}, \label{Eq_Tcc_fit_A}\\
   & V^B_\text{fit}(r;m_{\pi})=\sum\limits_{i=1,2}a_ie^{- \left({r}/{b_i}\right)^2} +a_3\left(1-e^{- \left({r}/{b_3}\right)^2}\right)^2 \frac{e^{-2m_\pi r}}{r^2}. \label{Eq_Tcc_fit_B}
\end{align}
As shown in Fig.~\ref{Fig-Tcc-V} by bands, two fit functions describe the lattice data equally well. Therefore, we use both to calculate physical observables and account their differences as a source of systematic error quoted in our final results.   

%==========================
%==========================

\begin{figure*}[htbp]
    \centering
    \includegraphics[width=7.0cm]{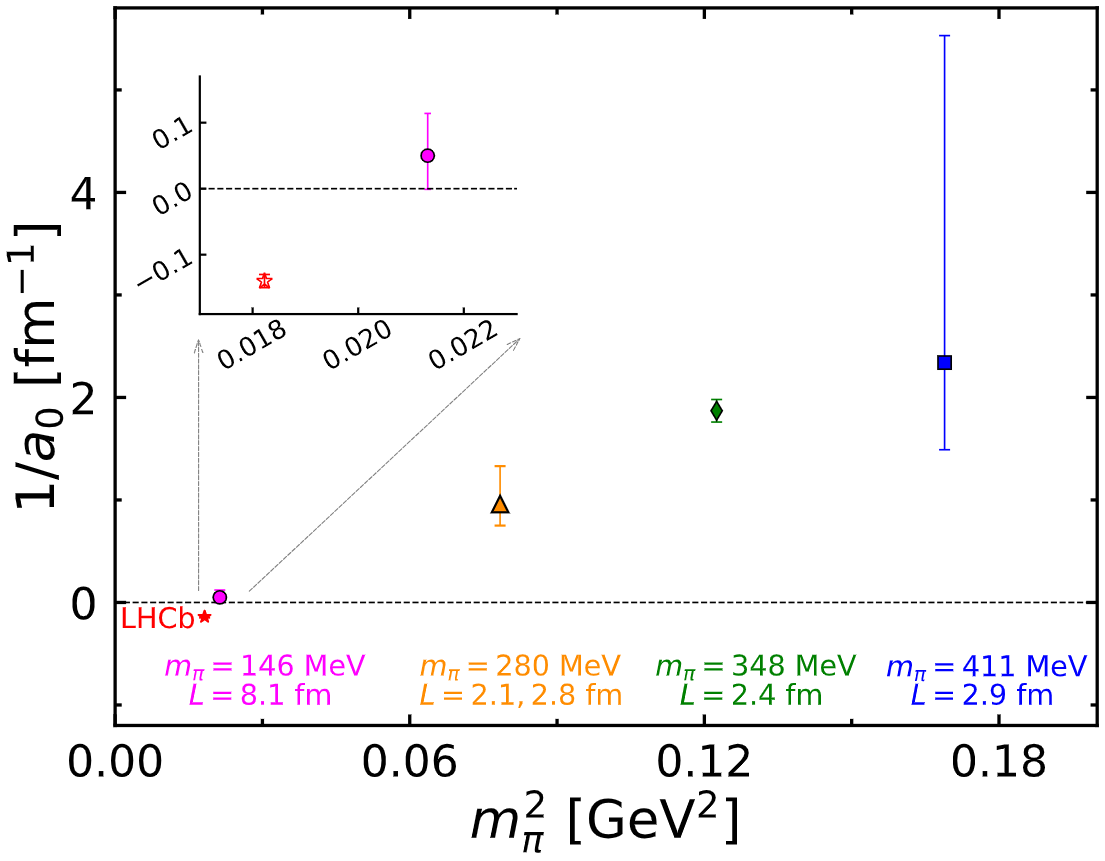}
    \includegraphics[width=7.4cm]{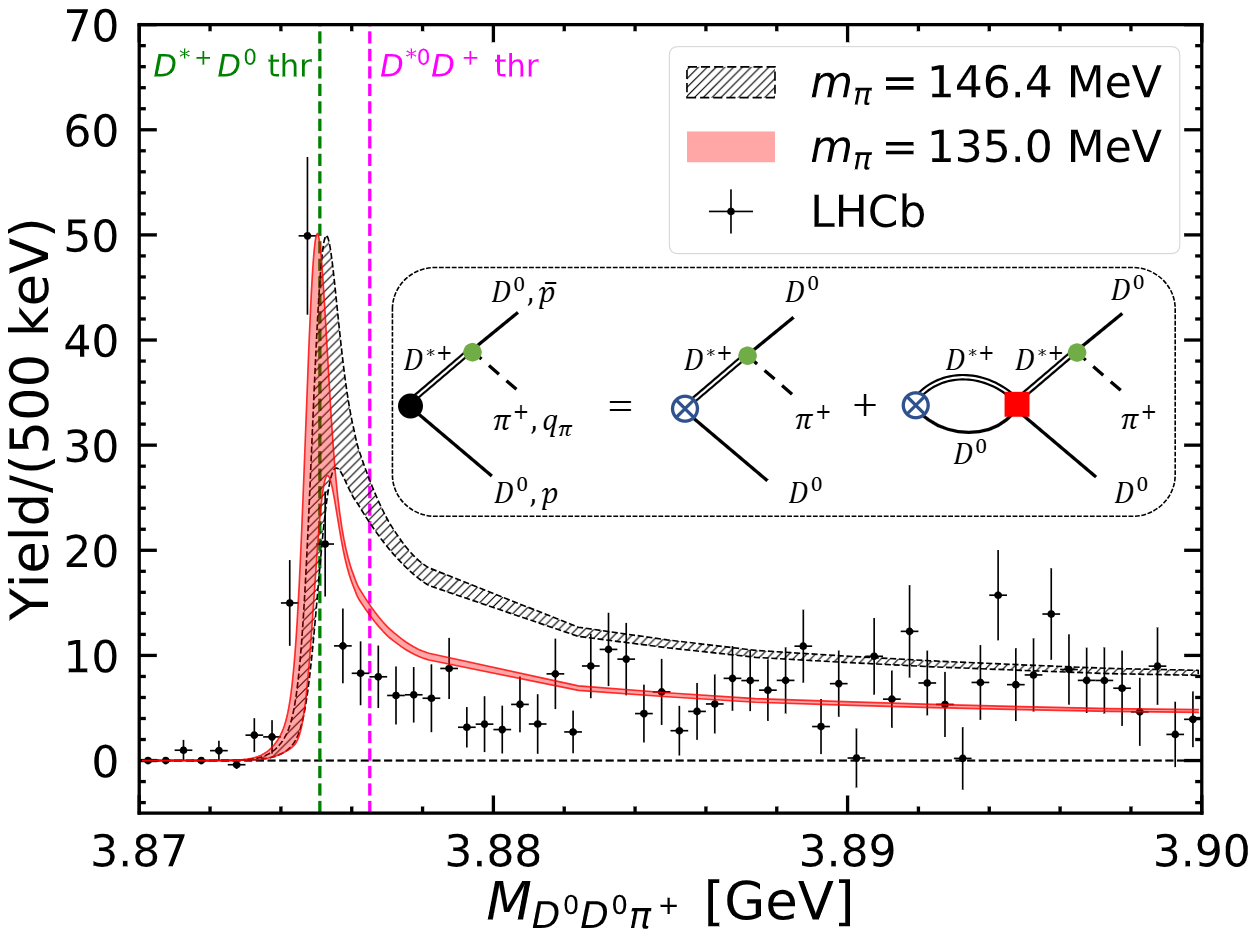}
    \caption{(Left) The inverse of scattering length $1/a_0$ for the $D^*D$ scattering in the $T^+_{cc}$ channel obtained from lattice QCD simulations by Refs.~\cite{Ikeda2013}(blue square),~\cite{Chen2022}(green diamond), and ~\cite{Padmanath2022}(yellow triangle).
 Our result~\cite{Lyu:2023xro}(magenta circle) and the real part of the experimental value by LHCb  (red star)~\cite{LHCb2021} are also shown.
 (Right) The $D^0D^0\pi^+$ mass spectrum calculated using $V^B_\text{fit}(r; m_\pi)$ with $m_\pi=146.4$ MeV ($m_\pi=135.0$ MeV) is shown by the black (red) band.
The black dots are LHCb data~\cite{LHCb2021}. 
The inset shows diagrams contributing to the $D^0D^0\pi^+$ mass spectrum, where the black filled circle, blue cross circle, green filled circle, 
and red square denote the production amplitude $U$, constant vertex $P$, $D^{*+}\rightarrow D^0\pi^+$ vertex, and $D^*D$ scattering $T$ matrix, respectively.
    }
    \label{Fig-Tcc-ao}
\end{figure*}

Shown in Table \ref{tab-scattering} are the scattering parameters as well as the pole positions $\kappa_\text{pole}$ \footnote{ The pole position in complex momentum plane is given by $k_\text{pole}=i\kappa_\text{pole}$.}, and $E_\text{pole}$ of the scattering matrix calculated with the fitted potentials.
The results show that $T^+_{cc}$ appears as a near-threshold virtual state in current lattice pion mass.
A comparison between lattice results and the LHCb data is shown in Fig.~\ref{Fig-Tcc-ao} (left).
To estimate how the scattering parameters change and how the pole evolves toward the physical point, 
we extrapolate the lattice potential to the physical point by $V^B_\text{fit}(r;m_{\pi}=146.4\rightarrow135.0~\text{MeV})$.
Using such a potential together with the physical hadron masses, we found that  $T^+_{cc}$ evolves into a loosely bound state (see  Table \ref{tab-scattering}).
Note that, the pole positions in both cases lie above the branch point of the left-hand cut (lhc) arising from one-pion exchange, 
rendering them unaffected by potential effects of the cut. 
For a detailed discussion of the lhc in the context of $T^+_{cc}$, refer to a dedicated study in Ref.~\cite{Aoki:2025jvi}.

%==========================
\begin{table}[htbp]
\begin{center}
\caption{The scattering parameters $1/a_0$ and $r_{\rm eff}$, together with the pole positions $\kappa_\text{pole}$, and $E_\text{pole}$.
Numbers in the 2nd column with statistical error (first parenthesis) and systematic error (second parenthesis) are obtained from $V^{A,B}_{\rm fit}(r)$ with $t/a=21-23$ at $m_\pi=146.4$ MeV.
The 3rd column shows estimated values from $V^B_{\rm fit}(r;m_{\pi})$ with $t/a=22$ and $m_\pi=135.0$ MeV.}
\begin{tabular}{lcc}
  \hline\hline
    $m_\pi$ [MeV]~~~~~~~~~&$146.4$~~~~~~~~~&$135.0$\\
  \hline
$1/a_0$ [fm$^{-1}$] ~~~~~~~~~&$0.05(5)\left(^{+2}_{-2}\right)$ ~~~~~~~~~~&$-0.03(4)$ \\
%\hline
$r_{\rm eff}$ [fm] ~~~~~~~~~&$1.12(3)\left(^{+3}_{-8}\right)$ ~~~~~~~~~~&$1.12(3)$ \\
%\hline
$\kappa_\text{pole}$ [MeV] ~~~~~~~~~&$-8(8)\left(^{+3}_{-5}\right)$ ~~~~~~~~~~&$+5(8)$ \\
%\hline
$E_\text{pole}$ [keV] ~~~~~~~~~&$-59\left(^{+53}_{-99}\right)\left(^{+2}_{-67}\right)$~~~~~~~~~~&$-45\left(^{+41}_{-78}\right)$\\
 \hline\hline
\end{tabular} \label{tab-scattering}    
\end{center}
\end{table}
%==========================

To make a further connection to the LHCb experimental data, we construct the $D^0D^0\pi^+$ mass spectrum by calculating the diagrams shown in the inset in Fig.~\ref{Fig-Tcc-ao} (right),
where the $D^*D$ interaction comes into play through the corresponding $T$ matrix shown by the red square.
The obtained $D^0D^0\pi^+$ mass spectrum using the potentials $V^B_\text{fit}(r; m_\pi)$ with $m_\pi=146.4$ MeV and $m_\pi=135.0$ MeV are shown in Fig.~\ref{Fig-Tcc-ao} (right) by the black and red bands, respectively.
We adopt experimental values for $m_{D^{*+}, D^0, \pi^+}$ and $\Gamma_{D^{*+}}$ ~\cite{PDG2020} in both calculations in order to keep the same phase space with the experiment.
The obtained two theoretical results exhibit a peak around the $D^{*+}D^0$ threshold, and the peak position moves to the left as the pion mass decreases toward the physical point.
In particular, the red band obtained at the physical point with $V^B_\text{fit}(r; m_\pi=135.0~\text{MeV})$ provides a semi-quantitative description to the LHCb data shown by the black dots.

%==========================
%==========================

\section{The $N$-$c\bar c$ interaction}
The low-energy interaction between a nucleon ($N$) and a charmonium ($c\bar c$) has intimate relations to  fundamental questions in QCD, 
such as how hadrons gain/lose mass in vacuum/medium, what kind of multiquark hadrons can exit.
In particular, a precise understanding of the $N$-$c\bar{c}$ interaction is essential for investigating
the trace anomaly contribution to the nucleon mass~\cite{Kharzeev:1995ij}, the modification of the $J/\psi$ mass in nuclear matter~\cite{Hayashigaki:1998ey},
the properties of the $P_c$ pentaquark states \cite{LHCb2015_Pc}, and the possible formation of charmonium-nucleus bound states~\cite{Krein:2017usp}.
However, the low-energy $N$-$c\bar{c}$ interaction has not been well constrained experimentally.
Recent measurements yield scattering lengths that differ by orders of magnitude:
one study using vector meson dominance reports values on the order of ${O}(1\sim10) \times 10^{-3}$~fm~\cite{Pentchev:2020kao},
while another based on low-energy unitarity finds values around ${O}(1)$~fm~\cite{JPAC:2023qgg}.
Previous lattice QCD studies~\cite{Yokokawa:2006td, Kawanai:2010ru, Skerbis:2018lew}
have been limited by quenched approximations, unphysically heavy pion masses, or large statistical uncertainties.
To address these limitations, Ref.~\cite{Lyu:2024ttm} has conducted a realistic lattice QCD calculation of the low-energy $N$-$c\bar{c}$ interaction.

\begin{figure*}[htbp]
    \centering
    \includegraphics[width=7.0cm]{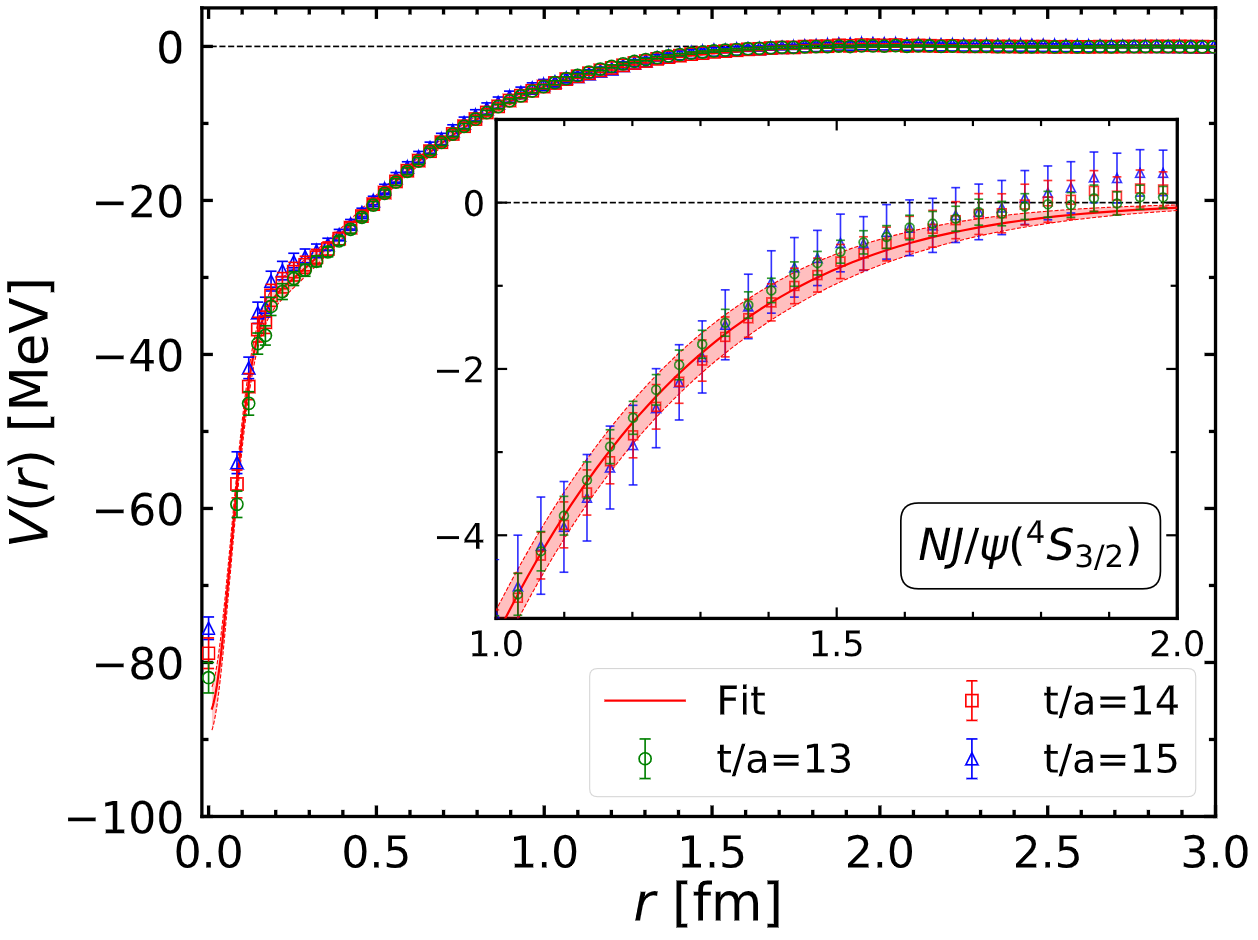}
    \includegraphics[width=7.0cm]{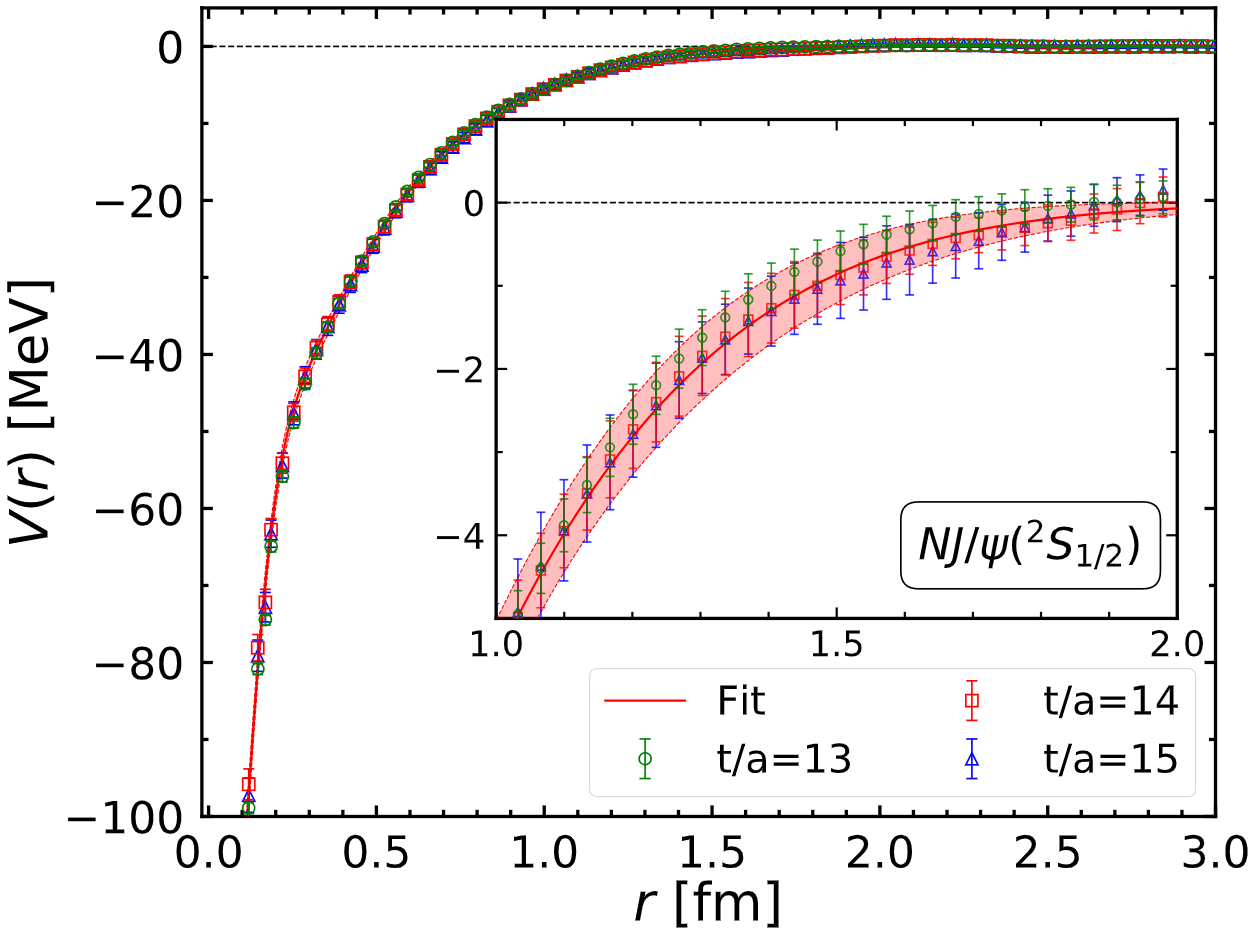}
    \includegraphics[width=7.0cm]{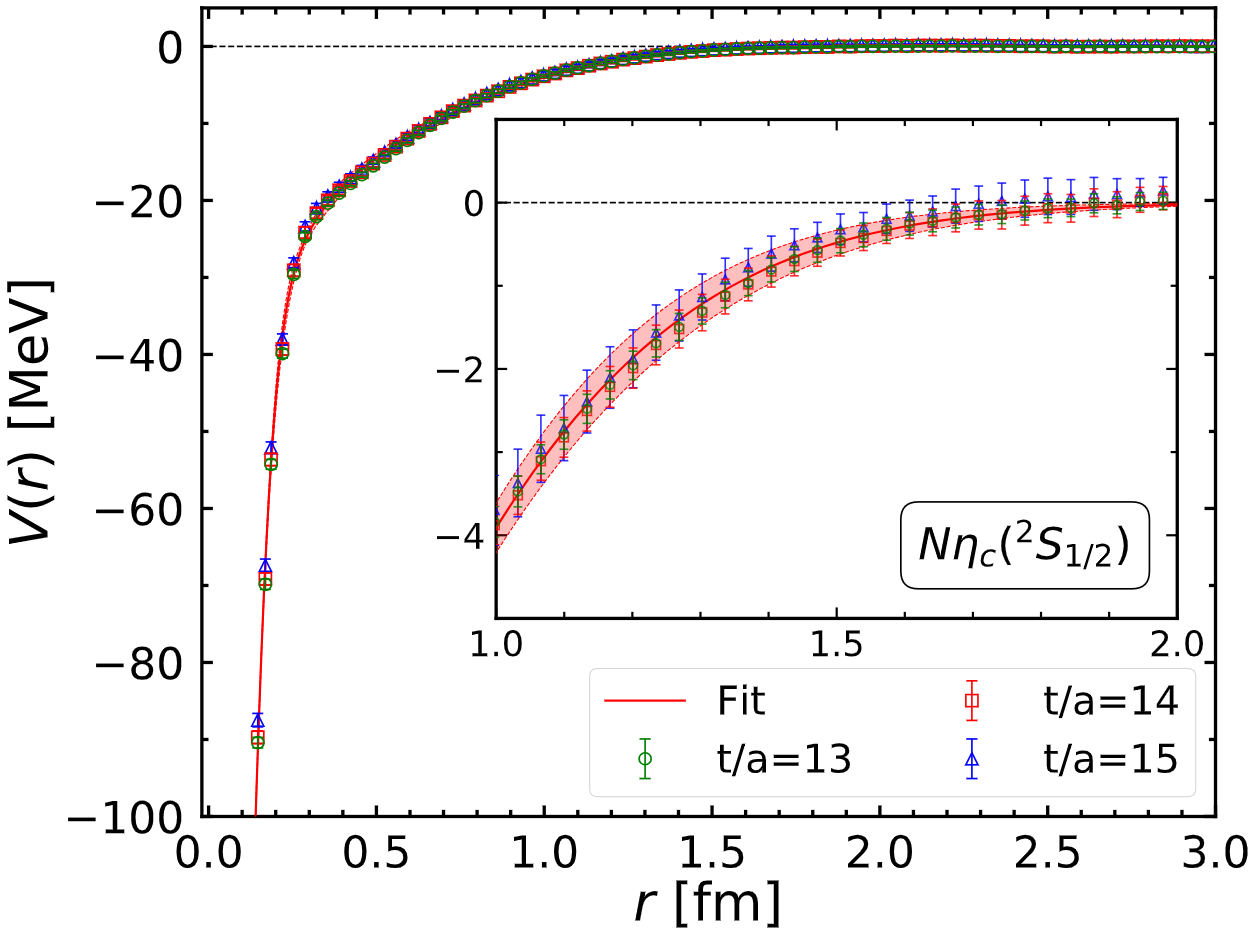}
    \includegraphics[width=7.0cm]{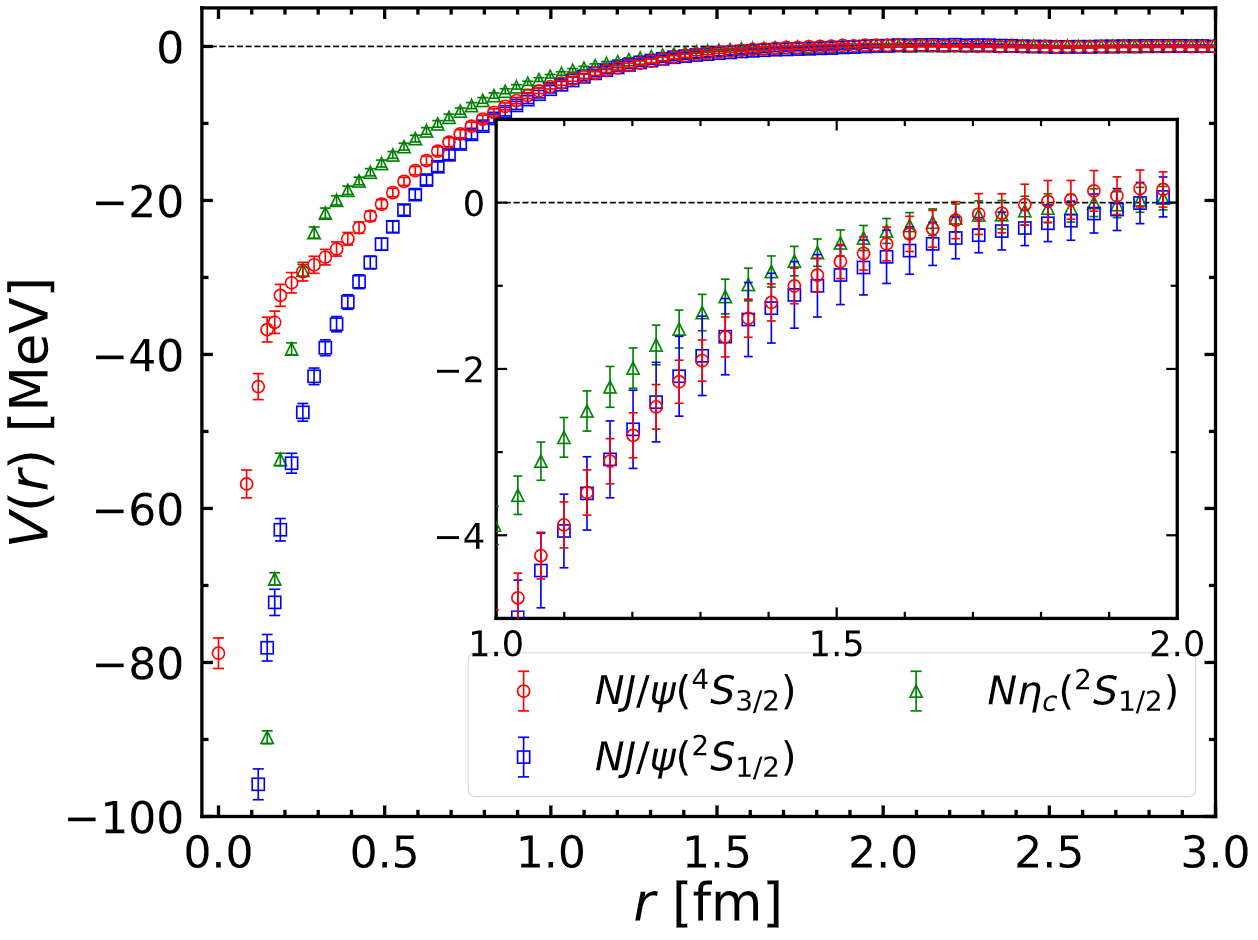}
    \caption{The $N$-$c\bar c$ potential in the $S$ wave for $N$-$J/\psi$ with spin $3/2$ (upper left), with spin $1/2$ (upper right), and for $N$-$\eta_c$ (lower left). 
    The red bands show the fitted results with a phenomenological three-range Gaussian at $t/a=14$.
    The three potentials at $t/a=14$ are also shown (lower right) for a direct comparison.
    A magnification is shown in the inset for each panel.
    }
    \label{Fig-Ncc-V}
\end{figure*}

\begin{figure*}[htbp]
    \centering
    \includegraphics[width=7.0cm]{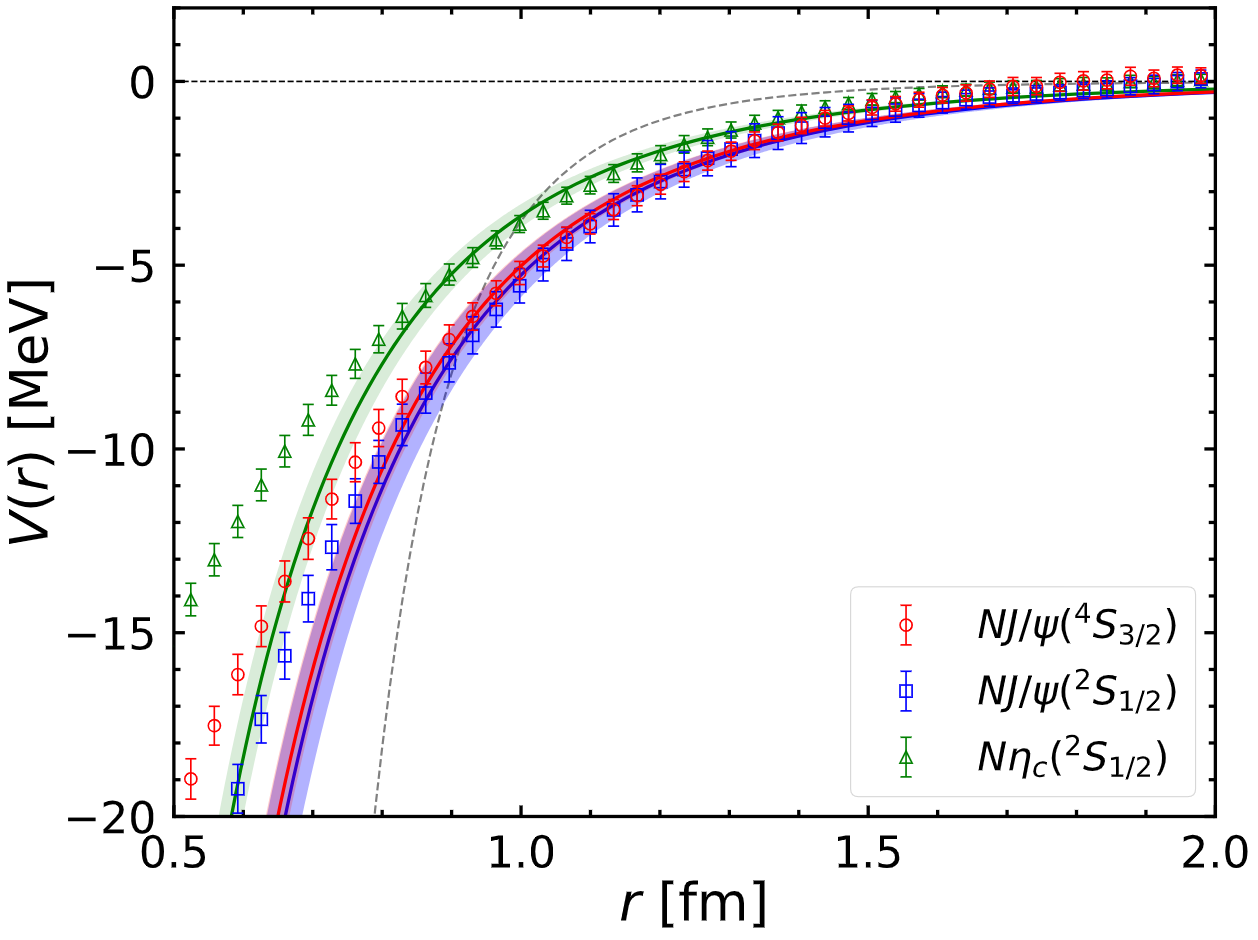}
    \includegraphics[width=7.0cm]{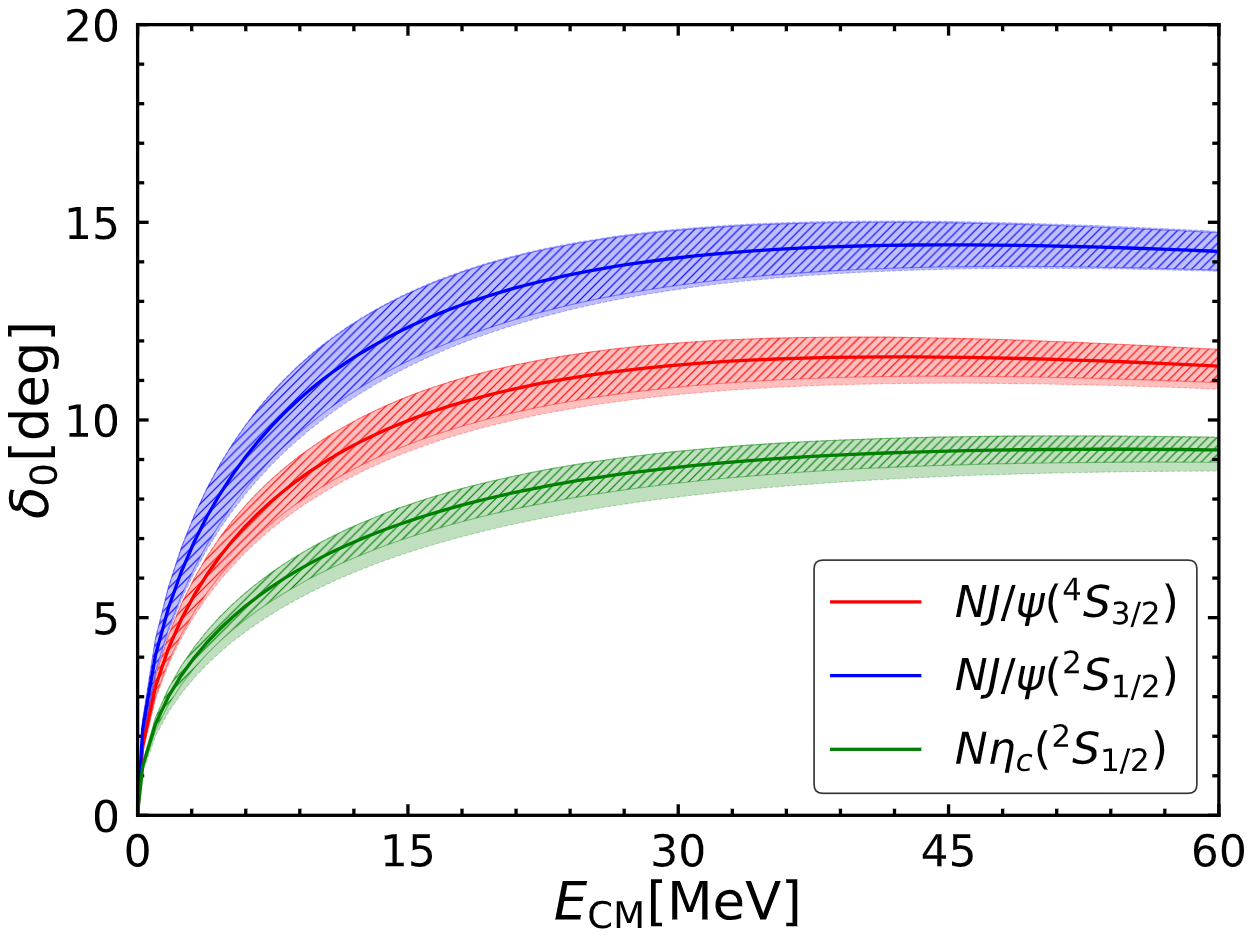}
    \caption{(Left) The TPE function $V(r)=-\alpha e^{-2m_\pi r}/r^2$ fitted to the long-range $N$-$c\bar{c}$ potentials. The gray dashed line is the best fit with $V(r)=-\alpha/r^7$ for comparison.
    (Right) The $N$-$c\bar c$ scattering phase shifts.
    }
    \label{Fig-Ncc-delta}
\end{figure*}

In Fig.~\ref{Fig-Ncc-V}, we show the $N$-$c\bar c$ potential in the $S$ wave for $N$-$J/\psi$ with spin $3/2$, with spin $1/2$, and for $N$-$\eta_c$.
The potentials in three channels are attractive at all distances, similar to the $N$-$s \bar s$ potential in Ref.~\cite{Lyu_Nphi_PRD2022}.
The similarity of the long-range potentials among these three channels shown in Fig.~\ref{Fig-Ncc-V} (lower right) indicates a common mechanism governs them.
As discussed in Refs~\cite{Fujii:1999xn,Castella2018}, the $N$-$c\bar c$ interaction generated through multiple gluon exchange is expected to 
manifest as the two-pion exchange interaction at long distances, which takes the form of $V(r)=-\alpha\frac{e^{-2m_\pi r}}{r^2}$ in the coordinate space.
Theoretically, such a potential is a QCD analogy of the van der Waals potential generated by two-photon exchange, which takes the form of  $V(r)=-\alpha/r^7$.
Indeed, we found  $V(r)=-\alpha\frac{e^{-2m_\pi r}}{r^2}$ is able to well describe lattice data at long distances, as shown in Fig. \ref{Fig-Ncc-delta} (Left).

%==========================
%==========================

In Fig.~\ref{Fig-Ncc-delta} (right), we  show the scattering phase shifts  calculated by solving the Schr\"odinger equation in the infinite volume with the potentials in Fig.~\ref{Fig-Ncc-V}.
The scattering parameters are summarized in Table~\ref{tab-scattering-Ncc}. 
These results indicate a mass reduction of $J/\psi$ in normal nuclear medium of density $\rho_\text{nm}=0.17~\text{fm}^{-3}$: $\delta m_{J/\psi}\simeq \frac{2\pi(m_N+ m_{J/\psi})}{m_Nm_{J/\psi}}a^\text{spin-av}_{J/\psi} \rho_\text{nm}=19(3)~\text{MeV}$
with $a^\text{spin-av}_{J/\psi}$ being the spin-average scattering length~\cite{Hayashigaki:1998ey}.

%==========================
%==========================

%==========================
\begin{table}[htbp]
\begin{center}
\caption{The $N$-$c\bar c$ scattering length $a_0$ and effective range $r_\text{eff}$.}
\begin{tabular}{lll}
  \hline\hline
    ~~~channel~~~~~~~~&~~~$a_0$ [fm]~~~~~~~~~&~~~$r_\text{eff}$ [fm]\\
  \hline
$N$-$J/\psi(^4S_{3/2})$ ~~~~~~~~~&$0.30(2)\left(^{+0}_{-2}\right)$ ~~~~~~~~~~&$3.25(12)\left(^{+6}_{-9}\right)$ \\
$N$-$J/\psi(^2S_{1/2})$ ~~~~~~~~~&$0.38(4)\left(^{+0}_{-3}\right)$ ~~~~~~~~~~&$2.66(21)\left(^{+0}_{-10}\right)$\\
$N\eta_c(^2S_{1/2})$ ~~~~~~~~~&$0.21(2)\left(^{+0}_{-1}\right)$ ~~~~~~~~~~&$3.65(19)\left(^{+0}_{-6}\right)$ \\
\hline\hline
\end{tabular} \label{tab-scattering-Ncc}
\end{center}
\end{table}   
%==========================

\section{The $\Omega_{ccc}$-$\Omega_{ccc}$ interaction}
The deuteron, composed of a proton and a neutron, is the only known stable dibaryon. 
In principle, QCD allows for the existence of other dibaryons. 
This has motivated a broad and ongoing search for other dibaryon states across different quark sectors.
Recent lattice QCD predictions of the $p\Omega$ dibaryon~\cite{Iritani2019PLB} and the $\Omega\Omega$ dibaryon~\cite{Gongyo2018} have renewed interest in  exploring whether stable dibaryons can also exist in the charm quark sector. 
In this context, Ref.~\cite{Lyu2021} investigated the simplest charmed dibaryon system, the $\Omega_{ccc}$-$\Omega_{ccc}$ in the $^1S_0$ channel, to study the possibility of binding.

\begin{figure*}[htbp]
    \centering
    \includegraphics[width=7.0cm]{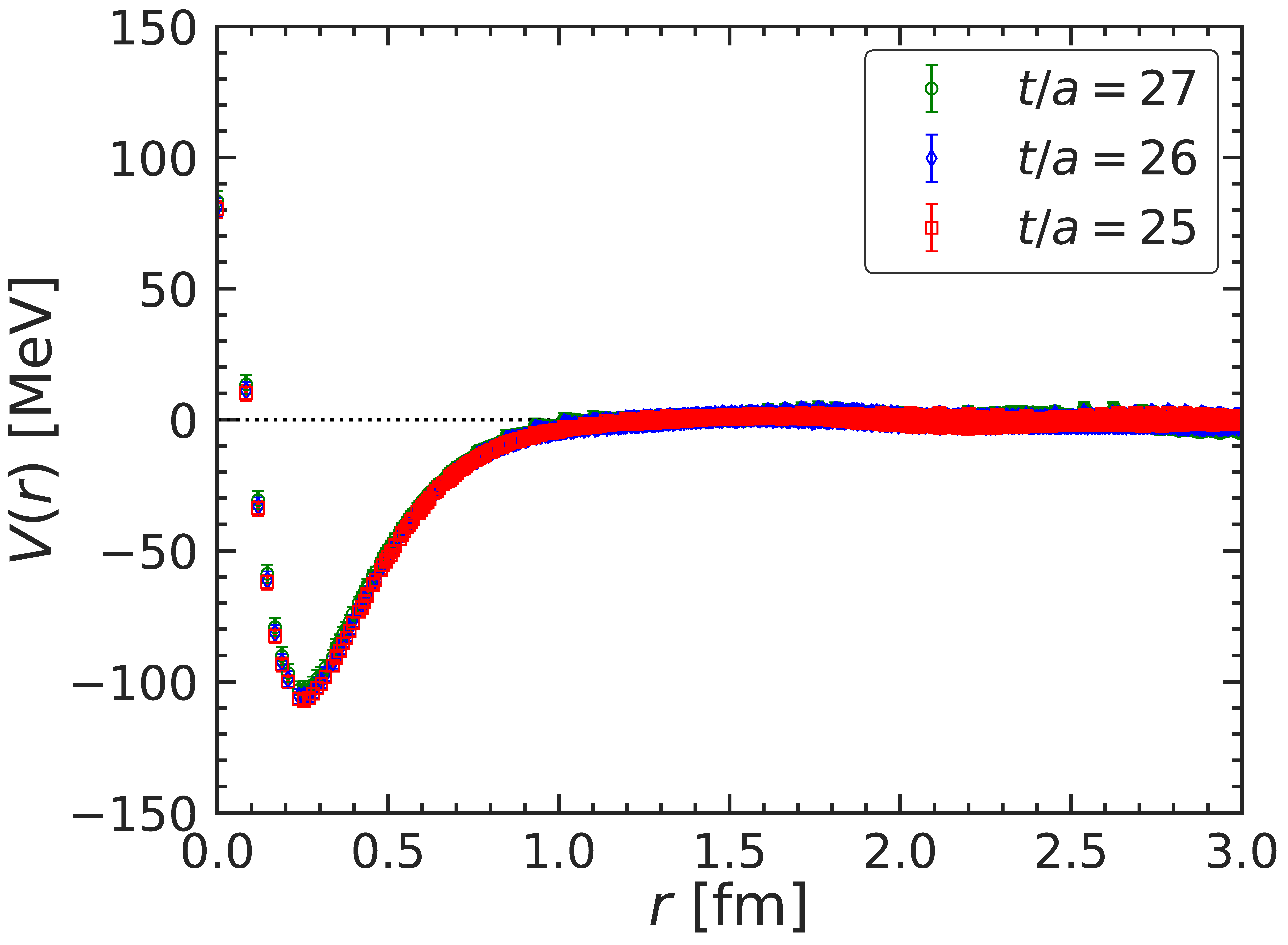}
    \includegraphics[width=7.0cm]{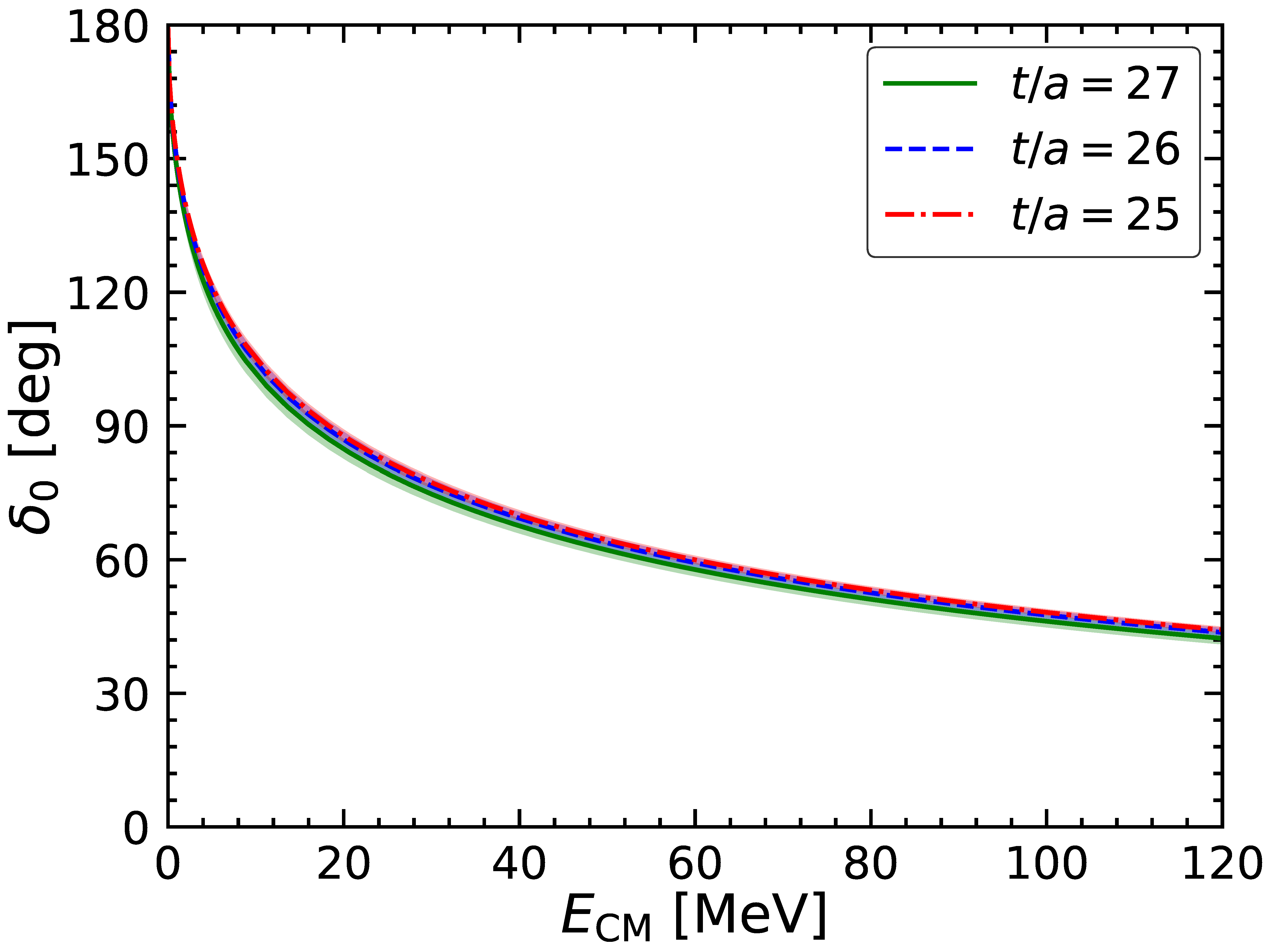}
    \caption{(Left) The $\Omega_{ccc}$-$\Omega_{ccc}$ potential in the $^1S_0$ channel.
    (Right) The $\Omega_{ccc}$-$\Omega_{ccc}$ scattering phase shifts in the $^1S_0$ channel.
    }
    \label{Fig-Omegaccc}
\end{figure*}

In Fig.~\ref{Fig-Omegaccc} (left), we show the $\Omega_{ccc}$-$\Omega_{ccc}$ interaction in the $^1S_0$ channel, which exhibits a short-range repulsive core surrounded by an attractive well at mid range, similar to the nuclear force and the
$\Omega\Omega$ potential in Ref.~\cite{Gongyo2018}.
The repulsive core at short distances may be qualitatively explained by the color-magnetic interaction between quarks in the phenomenological quark model~\cite{Oka1987},
which is expected to be proportional to $1/m^2_q$ with $m_q$ being the mass of the constituent quark $q$. 
We found that the potential at origin for $\Omega_{ccc}$-$\Omega_{ccc}$ is ten times weaker than that for $\Omega$-$\Omega$, which is qualitatively consistent with $m^2_s/m^2_c\sim 0.1$.
To convert the potential to physical observables, we fit the potential using a phenomenological three-range  Gaussian.
The scattering phase shifts calculated with the potential are shown in Fig.~\ref{Fig-Omegaccc} (right), which indicate the existence of a bound $\Omega_{ccc}$-$\Omega_{ccc}$ state in the $^1S_0$ channel.
The scattering parameters, binding energy, and the root-mean-square distance of the bound state are,
\begin{align}
       & a_0=-1.57(0.08)(_{-0.12}^{+0.04})~\text{fm}, \qquad r_{\mathrm{eff}}=0.57(0.02)(^{+0.01}_{-0.00})~\text{fm}, \\
       &B=5.68(0.77)(^{+0.46}_{-1.02})~\text{MeV},  \qquad  \sqrt{\langle r^2\rangle}=1.13(0.06)(^{+0.08}_{-0.03})~\text{fm}.
\end{align}

To investigate the effect from the Coulomb repulsion between two $\Omega_{ccc}$ baryons, we consider the following Coulomb potential taking into account the finite size of the $\Omega_{ccc}$ baryon,
\begin{align}
    V^{\text{Coulomb}}(r)= \frac{4\alpha_e}{r}\left[1 - e^{-x} \left(1 + \frac{11}{16} x + \frac{3}{16} x^2 + \frac{1}{48} x^3 \right) \right],
\end{align}
where $ x=2 \sqrt{6}r/r_d$ with $r_d=0.410(6)~\mathrm{fm}$~\cite{Can2015} being the charge radius of $\Omega_{ccc}$.
Using such a Coulomb potential together with the potential in Fig~\ref{Fig-Omegaccc} (left), we found that $\Omega_{ccc}$-$\Omega_{ccc}$ becomes unbound, but is located near the unitarity with scattering parameters shown as follows.
\begin{align}
    a^{\rm C}_0=19(7)(_{-7}^{+6})~\text{fm},\qquad r^{\rm C}_{\mathrm{eff}}=0.45(0.01)(^{+0.01}_{-0.00})~\text{fm}.
\end{align}
We also note that the ratio $r^{\rm C}_{\rm eff}/a^{\rm C}_0\simeq 0.024$ is considerably smaller in magnitude than that of the dineutron  ($0.149$).

%==========================
%==========================

\section{Summary}

In summary, we have reviewed recent progress on lattice QCD studies of charmed hadron interactions. 
Special attention has been paid to three representative channels 
investigated using lattice configurations with nearly physical quark masses ($m_\pi \simeq 146$ MeV) within the HAL QCD approach:
\begin{enumerate}
    \item[(i)] The $D^*$-$D$ interaction is sufficiently strong to produce a near-threshold virtual state, which evolves into a loosely bound state as $m_\pi$ approaches its physical value. 
        The $DD\pi$ mass spectrum computed from the potential offers a semi-quantitative description of the LHCb data.
     \item[(ii)] The $N$-$c\bar{c}$ interaction is attractive at all distances and exhibits a long-range two-pion exchange tail. 
        The obtained scattering parameters, as the first results derived under realistic conditions with controlled systematics, 
        not only offer crucial inputs for related studies but also provide key constraints to help resolve the tension between different experimental results.
    \item[(iii)] The $\Omega_{ccc}$-$\Omega_{ccc}$ potential supports a bound state in the $^1S_0$ channel. When Coulomb repulsion is included, the system is found to lie in the unitary regime.
\end{enumerate}

Lattice QCD plays a crucial role in elucidating charmed hadron interactions. 
The results presented here demonstrate its ability to connect nonperturbative dynamics of low-energy QCD with phenomena involving multi hadrons in the charm quark sector observed in experiments.
To advance our current understanding further, studies at the physical pion mass will be essential. Such efforts are currently underway.

\section*{Acknowledgements}

The author gratefully thanks members of the HAL QCD Collaboration for stimulating discussions. 
This work was partially supported by RIKEN Incentive Research Project (``Unveiling pion-exchange interactions between hadrons from first-principles lattice QCD''),  
the JSPS (Grant No. JP23H05439), 
and the Japan Science and Technology Agency (JST) as part of Adopting Sustainable Partnerships for Innovative Research Ecosystem (ASPIRE Grant No. JPMJAP2318).

%\bibliographystyle{JHEP}
%\bibliography{Reference}

\begin{thebibliography}{10}

\bibitem{Chen2022_PPP}
H.-X.~Chen, W.~Chen, X.~Liu, Y.-R.~Liu and S.-L.~Zhu, \emph{{An updated review of the new hadron states}}, \href{https://doi.org/10.1088/1361-6633/aca3b6}{\emph{Rept. Prog. Phys.} {\bfseries 86} (2023) 026201} [\href{https://arxiv.org/abs/2204.02649}{{\ttfamily 2204.02649}}].

\bibitem{Guo2017}
F.-K.~Guo, C.~Hanhart, U.-G.~Mei\ss{}ner, Q.~Wang, Q.~Zhao and B.-S.~Zou, \emph{{Hadronic molecules}}, \href{https://doi.org/10.1103/RevModPhys.90.015004}{\emph{Rev. Mod. Phys.} {\bfseries 90} (2018) 015004} [\href{https://arxiv.org/abs/1705.00141}{{\ttfamily 1705.00141}}].

\bibitem{Shen:2019dls}
S.~Shen, H.~Liang, W.H.~Long, J.~Meng and P.~Ring, \emph{{Towards an $ab initio$ covariant density functional theory for nuclear structure}}, \href{https://doi.org/10.1016/j.ppnp.2019.103713}{\emph{Prog. Part. Nucl. Phys.} {\bfseries 109} (2019) 103713} [\href{https://arxiv.org/abs/1904.04977}{{\ttfamily 1904.04977}}].

\bibitem{Hayano-Hatsuda2010}
R.S.~Hayano and T.~Hatsuda, \emph{Hadron properties in the nuclear medium}, \href{https://doi.org/10.1103/RevModPhys.82.2949}{\emph{Rev. Mod. Phys.} {\bfseries 82} (2010) 2949}.

\bibitem{Baym:2017whm}
G.~Baym, T.~Hatsuda, T.~Kojo, P.D.~Powell, Y.~Song and T.~Takatsuka, \emph{{From hadrons to quarks in neutron stars: a review}}, \href{https://doi.org/10.1088/1361-6633/aaae14}{\emph{Rept. Prog. Phys.} {\bfseries 81} (2018) 056902} [\href{https://arxiv.org/abs/1707.04966}{{\ttfamily 1707.04966}}].

\bibitem{Wiringa1995}
R.B.~Wiringa, V.G.J.~Stoks and R.~Schiavilla, \emph{Accurate nucleon-nucleon potential with charge-independence breaking}, \href{https://doi.org/10.1103/PhysRevC.51.38}{\emph{Phys. Rev. C} {\bfseries 51} (1995) 38}.

\bibitem{Machleidt:2000ge}
R.~Machleidt, \emph{{The High precision, charge dependent Bonn nucleon-nucleon potential (CD-Bonn)}}, \href{https://doi.org/10.1103/PhysRevC.63.024001}{\emph{Phys. Rev. C} {\bfseries 63} (2001) 024001} [\href{https://arxiv.org/abs/nucl-th/0006014}{{\ttfamily nucl-th/0006014}}].

\bibitem{Epelbaum:2014sza}
E.~Epelbaum, H.~Krebs and U.G.~Mei{\ss}ner, \emph{{Precision nucleon-nucleon potential at fifth order in the chiral expansion}}, \href{https://doi.org/10.1103/PhysRevLett.115.122301}{\emph{Phys. Rev. Lett.} {\bfseries 115} (2015) 122301} [\href{https://arxiv.org/abs/1412.4623}{{\ttfamily 1412.4623}}].

\bibitem{Entem:2017gor}
D.R.~Entem, R.~Machleidt and Y.~Nosyk, \emph{{High-quality two-nucleon potentials up to fifth order of the chiral expansion}}, \href{https://doi.org/10.1103/PhysRevC.96.024004}{\emph{Phys. Rev. C} {\bfseries 96} (2017) 024004} [\href{https://arxiv.org/abs/1703.05454}{{\ttfamily 1703.05454}}].

\bibitem{Lu:2021gsb}
J.-X.~Lu, C.-X.~Wang, Y.~Xiao, L.-S.~Geng, J.~Meng and P.~Ring, \emph{{Accurate Relativistic Chiral Nucleon-Nucleon Interaction up to Next-to-Next-to-Leading Order}}, \href{https://doi.org/10.1103/PhysRevLett.128.142002}{\emph{Phys. Rev. Lett.} {\bfseries 128} (2022) 142002} [\href{https://arxiv.org/abs/2111.07766}{{\ttfamily 2111.07766}}].

\bibitem{Aoyama:2024cko}
{\scshape HAL QCD} collaboration, \emph{{Scale setting and hadronic properties in the light quark sector with (2+1)-flavor Wilson fermions at the physical point}}, \href{https://doi.org/10.1103/PhysRevD.110.094502}{\emph{Phys. Rev. D} {\bfseries 110} (2024) 094502} [\href{https://arxiv.org/abs/2406.16665}{{\ttfamily 2406.16665}}].

\bibitem{Luscher1991}
M.~Luscher, \emph{{Two particle states on a torus and their relation to the scattering matrix}}, \href{https://doi.org/10.1016/0550-3213(91)90366-6}{\emph{Nucl. Phys. B} {\bfseries 354} (1991) 531}.

\bibitem{Ishii2007}
N.~Ishii, S.~Aoki and T.~Hatsuda, \emph{{Nuclear Force from Lattice QCD}}, \href{https://doi.org/10.1103/PhysRevLett.99.022001}{\emph{Phys. Rev. Lett.} {\bfseries 99} (2007) 022001}.

\bibitem{Ishii2012}
{\scshape HAL QCD Collaboration} collaboration, \emph{{Hadron-hadron interactions from imaginary-time Nambu-Bethe-Salpeter wave function on the lattice}}, \href{https://doi.org/https://doi.org/10.1016/j.physletb.2012.04.076}{\emph{Physics Letters B} {\bfseries 712} (2012) 437 }.

\bibitem{Lyu:2025lnd}
Y.~Lyu, S.~Aoki, T.~Doi, T.~Hatsuda, K.~Murakami and T.~Sugiura, \emph{{Decoding Two-Particle States in QCD with Spatial Wavefunctions}},  \href{https://arxiv.org/abs/2507.09930}{{\ttfamily 2507.09930}}.

\bibitem{Lyu:2025ncq}
Y.~Lyu, S.~Aoki, T.~Doi, T.~Hatsuda, K.~Murakami and T.~Sugiura, \emph{{Wavefunction-based operator optimization for two-hadron systems in lattice QCD}},  \href{https://arxiv.org/abs/2507.09933}{{\ttfamily 2507.09933}}.

\bibitem{Ishikawa2016}
{\scshape PACS Collaboration} collaboration, \emph{{2+1 Flavor QCD Simulation on a $96^4$ Lattice}}, \href{https://doi.org/10.22323/1.251.0075}{\emph{\textit{Proc. Sci.}} {\bfseries LATTICE2015} (2016) 075} [\href{https://arxiv.org/abs/1511.09222}{{\ttfamily 1511.09222}}].

\bibitem{Namekawa2017}
{\scshape PACS Collaboration} collaboration, \emph{{Charm physics by $N_f=2+1$ Iwasaki gauge and the six stout smeared $O(a)$-improved Wilson quark actions on a $96^4$ lattice}}, \href{https://doi.org/10.22323/1.256.0125}{\emph{\textit{Proc. Sci.}} {\bfseries LATTICE2016} (2017) 125}.

\bibitem{PDG2020}
{\scshape {Particle Data Group}} collaboration, \emph{{Review of Particle Physics}}, {\emph{Progress of Theoretical and Experimental Physics} {\bfseries 2020} (2020) }.

\bibitem{LHCb2021_NP}
{\scshape LHCb} collaboration, \emph{{Observation of an exotic narrow doubly charmed tetraquark}}, \href{https://doi.org/10.1038/s41567-022-01614-y}{\emph{Nature Phys.} {\bfseries 18} (2022) 751} [\href{https://arxiv.org/abs/2109.01038}{{\ttfamily 2109.01038}}].

\bibitem{LHCb2021}
{\scshape LHCb} collaboration, \emph{{Study of the doubly charmed tetraquark $T_{cc}^{+}$}}, \href{https://doi.org/10.1038/s41467-022-30206-w}{\emph{Nature Commun.} {\bfseries 13} (2022) 3351} [\href{https://arxiv.org/abs/2109.01056}{{\ttfamily 2109.01056}}].

\bibitem{Lyu:2023xro}
Y.~Lyu, S.~Aoki, T.~Doi, T.~Hatsuda, Y.~Ikeda and J.~Meng, \emph{{Doubly Charmed Tetraquark Tcc+ from Lattice QCD near Physical Point}}, \href{https://doi.org/10.1103/PhysRevLett.131.161901}{\emph{Phys. Rev. Lett.} {\bfseries 131} (2023) 161901} [\href{https://arxiv.org/abs/2302.04505}{{\ttfamily 2302.04505}}].

\bibitem{Noh2021}
S.~Noh, W.~Park and S.H.~Lee, \emph{{The Doubly-heavy Tetraquarks ($qq'\bar{Q}\bar{Q'}$) in a Constituent Quark Model with a Complete Set of Harmonic Oscillator Bases}}, \href{https://doi.org/10.1103/PhysRevD.103.114009}{\emph{Phys. Rev. D} {\bfseries 103} (2021) 114009} [\href{https://arxiv.org/abs/2102.09614}{{\ttfamily 2102.09614}}].

\bibitem{Aoki:2025jvi}
S.~Aoki, T.~Doi and Y.~Lyu, \emph{{Left-hand cut and the HAL QCD method}}, \href{https://doi.org/10.22323/1.466.0089}{\emph{PoS} {\bfseries LATTICE2024} (2025) 089} [\href{https://arxiv.org/abs/2501.16804}{{\ttfamily 2501.16804}}].

\bibitem{Ikeda2013}
Y.~Ikeda, B.~Charron, S.~Aoki, T.~Doi, T.~Hatsuda, T.~Inoue et~al., \emph{{Charmed tetraquarks $T_{cc}$ and $T_{cs}$ from dynamical lattice QCD simulations}}, \href{https://doi.org/10.1016/j.physletb.2014.01.002}{\emph{Phys. Lett. B} {\bfseries 729} (2014) 85} [\href{https://arxiv.org/abs/1311.6214}{{\ttfamily 1311.6214}}].

\bibitem{Chen2022}
S.~Chen, C.~Shi, Y.~Chen, M.~Gong, Z.~Liu, W.~Sun et~al., \emph{{$T^+_{cc}(3875)$ relevant DD scattering from Nf$=$2 lattice QCD}}, \href{https://doi.org/10.1016/j.physletb.2022.137391}{\emph{Phys. Lett. B} {\bfseries 833} (2022) 137391} [\href{https://arxiv.org/abs/2206.06185}{{\ttfamily 2206.06185}}].

\bibitem{Padmanath2022}
M.~Padmanath and S.~Prelovsek, \emph{{Signature of a Doubly Charm Tetraquark Pole in DD* Scattering on the Lattice}}, \href{https://doi.org/10.1103/PhysRevLett.129.032002}{\emph{Phys. Rev. Lett.} {\bfseries 129} (2022) 032002} [\href{https://arxiv.org/abs/2202.10110}{{\ttfamily 2202.10110}}].

\bibitem{Kharzeev:1995ij}
D.~Kharzeev, \emph{{Quarkonium interactions in QCD}}, \href{https://doi.org/10.3254/978-1-61499-215-8-105}{\emph{Proc. Int. Sch. Phys. Fermi} {\bfseries 130} (1996) 105} [\href{https://arxiv.org/abs/nucl-th/9601029}{{\ttfamily nucl-th/9601029}}].

\bibitem{Hayashigaki:1998ey}
A.~Hayashigaki, \emph{{\ensuremath{J/\psi} nucleon scattering length and in-medium mass shift of \ensuremath{J/\psi} in QCD sum rule analysis}}, \href{https://doi.org/10.1143/PTP.101.923}{\emph{Prog. Theor. Phys.} {\bfseries 101} (1999) 923} [\href{https://arxiv.org/abs/nucl-th/9811092}{{\ttfamily nucl-th/9811092}}].

\bibitem{LHCb2015_Pc}
{\scshape LHCb Collaboration} collaboration, \emph{Observation of $j/\ensuremath{\psi}p$ resonances consistent with pentaquark states in ${\mathrm{\ensuremath{\Lambda}}}_{b}^{0}\ensuremath{\rightarrow}j/\ensuremath{\psi}{K}^{\ensuremath{-}}p$ decays}, \href{https://doi.org/10.1103/PhysRevLett.115.072001}{\emph{Phys. Rev. Lett.} {\bfseries 115} (2015) 072001}.

\bibitem{Krein:2017usp}
G.~Krein, A.W.~Thomas and K.~Tsushima, \emph{{Nuclear-bound quarkonia and heavy-flavor hadrons}}, \href{https://doi.org/10.1016/j.ppnp.2018.02.001}{\emph{Prog. Part. Nucl. Phys.} {\bfseries 100} (2018) 161} [\href{https://arxiv.org/abs/1706.02688}{{\ttfamily 1706.02688}}].

\bibitem{Pentchev:2020kao}
L.~Pentchev and I.I.~Strakovsky, \emph{{$J/\psi$-$p$ Scattering Length from the Total and Differential Photoproduction Cross Sections}}, \href{https://doi.org/10.1140/epja/s10050-021-00364-4}{\emph{Eur. Phys. J. A} {\bfseries 57} (2021) 56} [\href{https://arxiv.org/abs/2009.04502}{{\ttfamily 2009.04502}}].

\bibitem{JPAC:2023qgg}
{\scshape Joint Physics Analysis Center} collaboration, \emph{{Dynamics in near-threshold \ensuremath{J/\psi} photoproduction}}, \href{https://doi.org/10.1103/PhysRevD.108.054018}{\emph{Phys. Rev. D} {\bfseries 108} (2023) 054018} [\href{https://arxiv.org/abs/2305.01449}{{\ttfamily 2305.01449}}].

\bibitem{Yokokawa:2006td}
K.~Yokokawa, S.~Sasaki, T.~Hatsuda and A.~Hayashigaki, \emph{{First lattice study of low-energy charmonium-hadron interaction}}, \href{https://doi.org/10.1103/PhysRevD.74.034504}{\emph{Phys. Rev. D} {\bfseries 74} (2006) 034504} [\href{https://arxiv.org/abs/hep-lat/0605009}{{\ttfamily hep-lat/0605009}}].

\bibitem{Kawanai:2010ru}
T.~Kawanai and S.~Sasaki, \emph{{Charmonium-nucleon interaction from lattice QCD with a relativistic heavy quark action}}, \href{https://doi.org/10.22323/1.105.0156}{\emph{PoS} {\bfseries LATTICE2010} (2010) 156} [\href{https://arxiv.org/abs/1011.1322}{{\ttfamily 1011.1322}}].

\bibitem{Skerbis:2018lew}
U.~Skerbis and S.~Prelovsek, \emph{{Nucleon-$J/\psi$ and nucleon-$\eta_{c}$ scattering in $P_{c}$ pentaquark channels from LQCD}}, \href{https://doi.org/10.1103/PhysRevD.99.094505}{\emph{Phys. Rev. D} {\bfseries 99} (2019) 094505} [\href{https://arxiv.org/abs/1811.02285}{{\ttfamily 1811.02285}}].

\bibitem{Lyu:2024ttm}
Y.~Lyu, T.~Doi, T.~Hatsuda and T.~Sugiura, \emph{{Nucleon-charmonium interactions from lattice QCD}}, \href{https://doi.org/10.1016/j.physletb.2024.139178}{\emph{Phys. Lett. B} {\bfseries 860} (2025) 139178} [\href{https://arxiv.org/abs/2410.22755}{{\ttfamily 2410.22755}}].

\bibitem{Lyu_Nphi_PRD2022}
Y.~Lyu, T.~Doi, T.~Hatsuda, Y.~Ikeda, J.~Meng, K.~Sasaki et~al., \emph{{Attractive \ensuremath{N}-\ensuremath{\phi} interaction and two-pion tail from lattice QCD near physical point}}, \href{https://doi.org/10.1103/PhysRevD.106.074507}{\emph{Phys. Rev. D} {\bfseries 106} (2022) 074507}.

\bibitem{Fujii:1999xn}
H.~Fujii and D.~Kharzeev, \emph{{Long range forces of QCD}}, \href{https://doi.org/10.1103/PhysRevD.60.114039}{\emph{Phys. Rev. D} {\bfseries 60} (1999) 114039} [\href{https://arxiv.org/abs/hep-ph/9903495}{{\ttfamily hep-ph/9903495}}].

\bibitem{Castella2018}
J.~Tarr\'us~Castell\`a and G.a.~Krein, \emph{{Effective field theory for the nucleon-quarkonium interaction}}, \href{https://doi.org/10.1103/PhysRevD.98.014029}{\emph{Phys. Rev. D} {\bfseries 98} (2018) 014029} [\href{https://arxiv.org/abs/1803.05412}{{\ttfamily 1803.05412}}].

\bibitem{Iritani2019PLB}
{\scshape HAL QCD} collaboration, \emph{{$N\Omega$ dibaryon from lattice QCD near the physical point}}, \href{https://doi.org/10.1016/j.physletb.2019.03.050}{\emph{Phys. Lett. B} {\bfseries 792} (2019) 284} [\href{https://arxiv.org/abs/1810.03416}{{\ttfamily 1810.03416}}].

\bibitem{Gongyo2018}
{\scshape HAL QCD Collaboration} collaboration, \emph{Most strange dibaryon from lattice qcd}, \href{https://doi.org/10.1103/PhysRevLett.120.212001}{\emph{Phys. Rev. Lett.} {\bfseries 120} (2018) 212001}.

\bibitem{Lyu2021}
Y.~Lyu, H.~Tong, T.~Sugiura, S.~Aoki, T.~Doi, T.~Hatsuda et~al., \emph{{Dibaryon with Highest Charm Number near Unitarity from Lattice QCD}}, \href{https://doi.org/10.1103/PhysRevLett.127.072003}{\emph{Phys. Rev. Lett.} {\bfseries 127} (2021) 072003} [\href{https://arxiv.org/abs/2102.00181}{{\ttfamily 2102.00181}}].

\bibitem{Oka1987}
M.~Oka, K.~Shimizu and K.~Yazaki, \emph{Hyperon-nucleon and hyperon-hyperon interaction in a quark model}, \href{https://doi.org/https://doi.org/10.1016/0375-9474(87)90371-X}{\emph{Nuclear Physics A} {\bfseries 464} (1987) 700 }.

\bibitem{Can2015}
K.U.~Can, G.~Erkol, M.~Oka and T.T.~Takahashi, \emph{{Look inside charmed-strange baryons from lattice QCD}}, \href{https://doi.org/10.1103/PhysRevD.92.114515}{\emph{Phys. Rev. D} {\bfseries 92} (2015) 114515} [\href{https://arxiv.org/abs/1508.03048}{{\ttfamily 1508.03048}}].

\end{thebibliography}

\providecommand{\href}[2]{#2}\begingroup\raggedright\endgroup

\end{document}